\documentclass[a4paper,11pt]{article}
\usepackage{pos}

\title{Coupled $\mathcal{N}$ = 2 supersymmetric quantum systems: 
symmetries and supervariable approach}
\ShortTitle{Coupled $\mathcal{N}$ = 2 SUSY QM: Symmetries and supervariable approach}

\author[a,b]{Aditi Pradeep}
\author[a]{Anjali S}
\author[a]{Binu M Nair}
\author*[a]{Saurabh Gupta}

\affiliation[a]{Department of Physics, \\ National Institute of Technology Calicut,\\ 
Kozhikode - 673 601, Kerala, India}

\affiliation[b]{Department of Physics and Astronomy, \\ University of British Columbia,
Vancouver Campus, \\ 325 - 6224 Agricultural Road, Vancouver, BC V6T 1Z1, Canada}

\emailAdd{aditipradeep314@gmail.com}
\emailAdd{anjalisujatha28@gmail.com}
\emailAdd{binu5120@gmail.com}
\emailAdd{saurabh@nitc.ac.in}

\abstract{We consider specific examples of $\mathcal{N}$ = 2 supersymmetric quantum mechanical models and list out all the novel symmetries. In each case, we show the existence of two sets of discrete symmetries that correspond to the Hodge duality operator of differential geometry. Thus, we are able to provide a proof of the conjecture which endorses the existence of more than one discrete symmetry transformation as the analogue of Hodge duality operation. Finally, we extend our analysis to a more general case and derive on-shell nilpotent symmetries within the framework of supervariable approach.
}

\FullConference{%
  40th International Conference on High Energy Physics - ICHEP2020\\
  July 28 - August 6, 2020\\
  Prague, Czech Republic (virtual meeting)
}

\begin{document}
\maketitle

\section{Introduction}
Supersymmetry has for long been regarded as one of the most beautiful theories, 
well-crafted by both physics and mathematics assistance, that describes the universe. From particle physics and quantum mechanics \cite{1} to chaos theory and optics \cite{2}, supersymmetry has found applications in a wide range of fields. Non-relativistic supersymmetric (SUSY) quantum mechanics gives us an elegant factorization technique of the Hamiltonian that allows us to find the energy spectrum of the system without actually solving the Schrödinger equation (cf. \cite{1} for details).

As far as $\mathcal{N}$ = 2 SUSY quantum mechanical (QM) models are concerned it has been shown that the underlying symmetries of these models provide a physical realization of de Rham cohomological operators of differential geometry \cite{4, 5, 6, 7} at algebraic level. 
Furthermore the supervariable approach has been utilized, in the context of SUSY theories, to derive the off-shell nilpotent symmetries with the aid of (anti-)chiral  supervariables and SUSY invariant restrictions (SUSYIRs) \cite{10}.

In this write-up, we take a generic supersymmetric ``polynomial-type'' potential \cite{11} and provide an account of continuous and discrete symmetries. We exhibit a proof of the conjecture which states that there might exist more than one set of discrete symmetry transformations, for a supersymmetric system, that provide an analogue for the Hodge duality operator of differential geometry \cite{6}. Furthermore, supervariable approach has been employed to derive the on-shell symmetries of the coupled superpotential (cf. \cite{11} for detailed results that are discussed in this proceeding).

\section{Generalized superpotential: continuous and discrete symmetries}
We begin our investigation with the Lagrangian for generalized form of superpotential $W' = \displaystyle \sum_{j=a}^{b}\alpha_{j}x^{j}$, described by bosonic coordinate $(x)$ and fermionic variables ($\psi,\bar{\psi}$), as 
\begin{equation}
		L=\frac{\dot{x}^{2}}{2}-\frac{1}{2}\left(\sum_{j=a}^{b}\alpha_{j}x^{j}\right)^{2}+i\bar{\psi}\dot{\psi}
	-\left(\sum_{j=a}^{b}j\alpha_{j}x^{j-1}\right)\bar{\psi}\psi, \label{lg} \\
	\end{equation}
where prime represents the first order derivative with respect to the bosonic coordinate $x$
and overdot represents time derivative. The limits $a$ and $b$ could take any of the positive or negative integer values such that we obtain a polynomial type superpotential with 
$\alpha_{j}$'s  being the constant coefficients of each power of $x$. For example, let us choose the limit $a=-1$ and $b=1$ and identifying $\alpha_{-1} =\lambda$, $\alpha_{0} =\mu$ and $\alpha_{1}=\omega$, we obtain the form of generalized superpotential as $W'=\frac{\lambda}{x}+\mu +\omega x$. While substituting for $\lambda=0$, we retrieve superposition of harmonic oscillator superpotential with free particle and when $\mu=0$, we obtain the superposition of superpotential of harmonic oscillator with the Coulomb-like (cf. \cite{11} for details).

The Lagrangian \eqref{lg} is found to respect two continuous SUSY transformations $s_{1}$ and $s_{2}$, whose explicit 
forms are given as 
\begin{eqnarray}
s_{1}x =-i\psi, \quad s_{1}\psi =0, \quad s_{1}\bar{\psi} =\dot{x}+i\sum_{j=a}^{b}\alpha_{j}x^{j}, \label{s1g}\\
s_{2}x =i\bar{\psi},\quad s_{2}\bar{\psi} =0, \quad s_{2}\psi =-\dot{x}+i\sum_{j=a}^{b}\alpha_{j}x^{j}. \label{s2g}
\end{eqnarray}
Under these SUSY transformations $s_{1}$ and $s_{2}$, the Lagrangian \eqref{lg} 
transforms as
\begin{equation}
	s_{1}L =-\dfrac{d}{dt}\left(\sum_{j=a}^{b}\big(\alpha_{j}x^{j}\big)\psi\right), \qquad
	s_{2}L =\dfrac{d}{dt}\left(i\dot{x}\bar{\psi}\right).\\		\end{equation}
It is clear that $s_{1}$ and $s_{2}$ are symmetries of
Lagrangian \eqref{lg} as it transforms into a total time derivative under these continuous SUSY transformations. Another characteristic of these
continuous symmetry transformations is that, they are on-shell nilpotent of order two (i.e. $s_{1}^{2}$ = $s_{2}^{2}$ = 0). Moreover, one can obtain a bosonic symmetry transformation 
$s_{w}$ from the anti-commutator of $s_{1}$ and $s_{2}$ as $s_{w}=\left\lbrace s_{1},s_{2}\right\rbrace$. The explicit form of transformations of dynamical variables under $s_{w}$ is given as
\begin{equation}\label{swg}
		s_{w}x = 2i\dot{x}, \quad s_{w}\psi = i\dot{\psi}+\sum_{j=a}^{b}j\alpha_{j}x^{j-1}\psi,
	\quad s_{w}\bar{\psi} = i\dot{\bar{\psi}}-\sum_{j=a}^{b}j\alpha_{j}x^{j-1}\bar{\psi}.\\
	\end{equation}
The Lagrangian \eqref{lg} transforms, under the bosonic symmetry $s_{w}$, as
\begin{eqnarray}
	s_{w}L &=& i\dfrac{d}{dt}\left( \dot{x}^{2} - \big(\sum_{j=a}^{b}\alpha_{j}x^{j}\big)^{2}
	+ i\bar{\psi}\dot{\psi} - \big(\sum_{j=a}^{b}j\alpha_{j}x^{j-1}\big)\bar{\psi}\psi\right).
	\end{eqnarray}
Thus, our Lagrangian \eqref{lg} for generalized superpotential is found to respect two on-shell nilpotent fermionic symmetries (cf. \eqref{s1g}, \eqref{s2g}) and one bosonic symmetry (cf. \eqref{swg}).

Apart from the continuous symmetries present in the system, the Lagrangian \eqref{lg} also respect eight sets of discrete symmetry $(\ast)$ transformations (cf. \cite{11} for details). However, here we discuss only the two {\it relevant} sets of transformations as given below
\begin{eqnarray}  \label{ds2}
& x\longrightarrow \pm x, \; \quad t\longrightarrow -t, \;\quad \psi \longrightarrow  \pm \bar{\psi}, \;\quad \bar{\psi} \longrightarrow \mp \psi, \\
&\alpha_{2n+1} \longrightarrow \alpha_{2n+1}, \; \alpha_{2n} \longrightarrow \pm \alpha_{2n}, \; {\text {where  $n$ is any integer.}} & \nonumber
\end{eqnarray}
In the above, one set of discrete symmetry transformations endow parity symmetry and the other does not, along with the time-reversal symmetry.
Now, from the discrete symmetries listed in \eqref{ds2}, 
concentrating on the all upper signatures only (i.e. $x\longrightarrow + x, \;  t\longrightarrow -t, \; \psi \longrightarrow  + \bar{\psi}, \; \bar{\psi} \longrightarrow - \psi, \alpha_{2n+1} \longrightarrow \alpha_{2n+1}, \; \alpha_{2n} \longrightarrow + \alpha_{2n}$) and the one with all lower signatures only (i.e. $x\longrightarrow - x, \;  t\longrightarrow -t, \; \psi \longrightarrow  - \bar{\psi}, \; \bar{\psi} \longrightarrow + \psi, \alpha_{2n+1} \longrightarrow \alpha_{2n+1}, \; \alpha_{2n} \longrightarrow - \alpha_{2n}$), we found that these discrete symmetries along with the continuous nilpotent symmetries $s_{1}$ and $s_{2}$ satisfy following relationships 
\begin{eqnarray}
    s_{1}\phi_{1}=+ \ast s_{2}\ast\phi_{1}, \; \qquad s_{2}\phi_{1}= - \ast s_{1}\ast\phi_{1},\nonumber \\
	s_{1}\phi_{2} = - \ast s_{2}\ast\phi_{2}, \; \qquad s_{2}\phi_{2} = + \ast s_{1}\ast\phi_{2}.
	\label{Hodge_1}
\end{eqnarray}
Here $\phi_{1}$ and $\phi_{2}$ are bosonic and fermionic variables, respectively. The above relationship \eqref{Hodge_1} has exact similar structure, at the algebraic level, as the one existing among the de Rham cohomological operators, (co-)exterior derivatives ($(\delta) d$) and Hodge duality operator ($\star$) of differential geometry (i.e. $\delta \; = \; \pm \; \star \; d \; \star$). Moreover, the continuous symmetries ($s_{1}, s_{2}, s_{w}$) also satisfy the algebraic structures similar to that satisfied by de Rham cohomological operators $(d, \delta, \Delta)$ of differential geometry, where $\Delta$ is the Laplacian operator (cf. \cite{11} for details). Thus, we conclude that, two sets of discrete symmetries draw a physical realization with Hodge duality operator ($\star$), which in turn prove the conjecture that there exists more than one set of discrete symmetry transformations \cite{6,11}. Also, we can identify the continuous symmetries $s_{1},  s_{2}$ analogous to $\delta, d$ and the bosonic symmetry $s_{w}$ corresponds to $\Delta$ of differential geometry 
at the algebraic level \cite{11}.

\section{Supervariable approach: on-shell nilpotent symmetries}
We now concentrate to derive the full set of on-shell nilpotent fermionic symmetries on (1,2)-dimensional supermanifold with the use of supervariable approach. The supermanifold is characterized by the bosonic variable $t$ and the Grassmann variables $\theta, \bar{\theta}$ ($\theta^{2} = 0, \bar{\theta}^{2} = 0,
\theta\bar{\theta} + \bar{\theta}\theta = 0$). To accomplish this goal, we identify the dynamical variables onto the (anti-)chiral super-submanifold and implement  SUSYIRs.

\subsection{Anti-chiral supervariable approach}
We focus on the (1,1)-dimensional anti-chiral super-submanifold characterized by $t$ and $\bar{\theta}$. The generalization of ordinary variables to the corresponding anti-chiral supervariables yields
 \begin{eqnarray} \label{ac_1}
  {\cal{X}}(t,\bar{\theta}) = x(t)+\bar{\theta}\Lambda(t), \quad 
   \Psi(t,\bar{\theta}) = \psi(t)+i\bar{\theta}\Omega_{1}(t), \quad
 \bar{\Psi}(t,\bar{\theta}) =\bar{\psi}(t)+ i\bar{\theta}\Omega_{2}(t), 
 \end{eqnarray}
where $\Lambda(t)$ is a fermionic secondary variable and $\Omega_{1}(t), \Omega_{2}(t)$ are bosonic secondary variables. In the first place, taking into consideration, the invariance of $\psi$ under $s_{1}$ yields
\begin{equation} \label{susy_ac_1}
    \Omega_{1}(t) = 0.
\end{equation}
It is straightforward to prove that $s_{1}\big(x\psi\big) =  0$ and $s_{1}\big(\dot{x}\dot{\psi}\big)  =  0$ and equating these invariant quantities $x\psi$ and $\dot{x}\dot{\psi}$ to their corresponding anti-chiral supervariables and making use of obtained value of $\Omega_{1}(t)$ (cf. \eqref{susy_ac_1}), we obtain 
 \begin{equation} \label{susy_ac_2}
     \Lambda(t)  =  -i\psi(t).
 \end{equation}
 Now, considering the combination: $ \Upsilon(t) \; \equiv \; \frac{1}{2}\dot{x}^{2}(t) + i \bar{\psi}(t)\dot{\psi}(t)
 -  \frac{1}{2}\Big(\displaystyle \sum_{j=a}^{b}\alpha_{j}x^{j} (t) \Big)^{2} \\
- \Big(\displaystyle \sum_{j=a}^{b} j\alpha_{j}x^{j-1} (t) \Big) \bar{\psi} (t) \psi (t)
+ i \Big(\displaystyle \sum_{j=a}^{b} \alpha_{j}x^{j} (t) \Big)\dot{x} (t)$ which is invariant under 
$s_{1}$ (i.e. $s_{1}\Upsilon(t)=0$). This leads to the SUSYIR,  
$\Gamma(t,\bar{\theta}) \; = \; \Upsilon(t),$
where $\Gamma(t,\bar{\theta})$ represents the generalization of $\Upsilon(t)$ on the anti-chiral super-submanifold. Equating both sides of the above SUSYIR and substituting the values from \eqref{susy_ac_1} and \eqref{susy_ac_2}, we obtain 
\begin{equation} \label{susy_ac_3}
    \Omega_{2}(t)  = -i \Big(\dot{x} (t) + i \sum_{j=a}^{b}\alpha_{j}x^{j} (t) \Big).
\end{equation} 
Plugging in the obtained values of $\Omega_{1}(t)$,  $\Lambda(t)$ and $\Omega_{2}(t)$ into \eqref{ac_1}, we get
\begin{eqnarray} \label{gac_1}
&{\cal{X}}(t,\bar{\theta}) = x(t)+\bar{\theta}\big(-i\psi\big)
\equiv x(t)+\bar{\theta}\big(s_{1}x\big), \quad
\Psi(t,\bar{\theta}) = \psi(t)+\bar{\theta}\big(0\big)
\equiv \psi(t)+\bar{\theta}\big(s_{1}\psi\big),
&\\ \nonumber
&\bar{\Psi}(t,\bar{\theta}) = \bar{\psi}(t)+\bar{\theta}\Big(\dot{x}+i\displaystyle\sum_{j=a}^{b}\alpha_{j}x^{j}\Big)
\equiv \bar{\psi}(t)+\bar{\theta}\big(s_{1}\bar{\psi}\big).&
\end{eqnarray}
Thus, we have derived the on-shell nilpotent SUSY transformation $s_{1}$ with the aid of SUSYIRs on anti-chiral 
super-submanifold. Finally, we can establish the equivalence between the translational generator $(\partial_{\bar{\theta}})$ and $s_{1}$ as:
$\displaystyle \frac{\partial{}}{\partial{\bar{\theta}}}\Phi(t,\bar{\theta}) = s_{1}\phi(t)$, where $\Phi(t,\bar{\theta})$ is the general supervariable defined on the anti-chiral super-submanifold.

\subsection{Chiral supervariable approach}
In this section, we focus on the derivation of $s_{2}$ by making use of SUSYIRs on the (1,1)-dimensional chiral super-submanifold parameterized by ($t,\theta$). First of all, the generalization of ordinary variables to corresponding chiral supervariables yields
\begin{eqnarray} \label{c_1}
{\cal{X}}(t,\theta) = x(t)+\theta \tilde{\Lambda}(t), \quad
\Psi(t,\theta) = \psi(t)+i\theta \tilde{\Omega}_{1}(t), \quad
	 \bar{\Psi}(t,\theta) =\bar{\psi}(t)+ i{\theta}\tilde{\Omega}_{2}(t),
\end{eqnarray}
where $\tilde{\Lambda}(t)$ is fermionic secondary variable and $\tilde{\Omega}_{1}(t), \tilde{\Omega}_{2}(t)$ represent bosonic secondary variables. Since $\bar{\psi}(t)$ is invariant under $s_{2}$, we obtain
\begin{equation} \label{susy_c_1}
    \tilde{\Omega}_{2}(t) = 0.
\end{equation}
We note that the quantities $\big(x\bar{\psi}\big)$ and $\big(\dot{x}\dot{\bar{\psi}}\big)$ are invariant under $s_{2}$ and these invariant quantities with the aid of obtained value for $\tilde{\Omega}_{2}(t)$ (cf. \eqref{susy_c_1}), generate 
\begin{equation} \label{susy_c_2}
    \tilde{\Lambda}(t) = i\bar{\psi}(t).
\end{equation}
Further, we observe that the following quantity: $ \tilde{\Upsilon}(t) \equiv \frac{1}{2}\dot{x}^{2}(t)-i\dot{\bar{\psi}}(t){\psi}(t)
- \frac{1}{2}\Big(\displaystyle \sum_{j=a}^{b}\alpha_{j}x^{j} (t)\Big)^{2}\\ - \Big(\displaystyle \sum_{j=a}^{b} j\alpha_{j}x^{j-1} (t)\Big)
\bar{\psi} (t)\psi(t) - i\Big(\displaystyle \sum_{j=a}^{b} \alpha_{j}x^{j} (t) \Big)\dot{x} (t) $ remain invariant under $s_2$ (i.e. $s_2 \tilde{\Upsilon}(t) = 0$). Thus, we have SUSYIR,  $\tilde{\Gamma}(t,\theta) \; = \; \tilde{\Upsilon}(t),$ 
where $\tilde{\Gamma}(t,\theta)$ represents the supervariable counterpart of $\tilde{\Upsilon}(t)$. Plugging in values from \eqref{susy_c_1} and  \eqref{susy_c_2} into SUSYIR and equating both the sides, we get
\begin{equation} \label{susy_c_3}
    \tilde{\Omega}_{1}(t)=-i\Big(-\dot{x} (t) + i \sum_{j=a}^{b}\alpha_{j}x^{j} (t) \Big).
\end{equation}
 Substituting the obtained values of $\tilde{\Omega}_{2}(t)$, $\tilde{\Lambda}(t)$ and $\tilde{\Omega}_{1}(t)$ into \eqref{c_1}, we obtain
\begin{eqnarray} \label{gc}
&{\cal{X}}(t,\theta) = x(t)+\theta\big(i\bar{\psi}\big)
\equiv x(t)+\theta\big(s_{2}x\big), \quad
\bar{\Psi}(t,\theta) = \bar{\psi}(t)+\theta(0)
\equiv \bar{\psi}(t)+\theta\big(s_{2}\bar{\psi}\big),& \\ \nonumber
&\Psi(t,\theta) = \psi(t)+\theta\Big(-\dot{x}+i \displaystyle \sum_{j=a}^{b}\alpha_{j}x^{j}\Big) \equiv \psi(t)+\theta\big(s_{2}\psi\big).&
\end{eqnarray}
Therefore, on the chiral supersubmanifold we have derived the 
on-shell nilpotent SUSY transformation $s_{2}$ by using appropriate SUSYIRs. It is worth mentioning that, from inspecting \eqref{gc}, we can draw the equivalence $\displaystyle \frac{\partial}{\partial{\theta}}\Phi(t,\theta) = s_{2} \; \phi(t)$ where, $\frac{\partial}{\partial{\theta}}$ is the translational derivative along chiral direction and  $\Phi(t,\theta)$ is the general supervariable in the chiral super-submanifold. Before ending this subsection, it is worthwhile to mention that the nilpotent nature of translational generators proves the nilpotency of fermionic symmetries effortlessly.

\section{Conclusions}
In our present endeavor, we have illustrated that $\mathcal{N}$ = 2 SUSY QM model respects two fermionic on-shell nilpotent continuous symmetries and a bosonic symmetry. In addition, we have demonstrated that there are two sets of {\it relevant} discrete symmetries which leave the Lagrangian quasi-invariant. Along with the continuous symmetries it provides a physical realization of the de Rham cohomological operators at algebraic level. 
Thus, the existence of two sets of discrete symmetries provides a proof of the conjecture which endorses the existence of more than one set of discrete symmetry transformations as the analogue of Hodge duality operation.  

Furthermore, we have derived on-shell nilpotent fermionic symmetries for the $\mathcal{N}$ = 2 SUSY QM model with the help of supervariable approach. It is worthwhile to mention that the form of SUSYIRs utilized in our investigation gives a general form of invariant restrictions for any kind of ``polynomial-type'' superpotential falling under this category of study.  \\

\noindent
{\bf Acknowledgments} The support from FRG scheme of NIT Calicut is thankfully acknowledged.

\end{document}